\documentclass[journal]{IEEEtran}
\usepackage{amsmath,amssymb,cite,url}
\usepackage{amsthm}

\newtheoremstyle{mytheorem}
  {3pt}        
  {3pt}        
  {\itshape}   
  {}           
  {\bfseries}  
  {.}          
  {.5em}       
  {}           

\theoremstyle{mytheorem}
\newtheorem{theorem}{Theorem}
\title{Three Machines Suffice for Local Hardening of Electromechanical Oscillations}
\author{Kai~Sun,
\thanks{This work was supported by NSF grant ECCS-2329924.}%
\thanks{K. Sun is with the Department of EECS, The University of Tennessee, Knoxville, TN 37996 USA (e-mail: kaisun@utk.edu).}%
}
\begin{document}
\maketitle
\begin{abstract}
The single-degree-of-freedom swing equation with sinusoidal coupling is locally softening: its oscillation frequency decreases with oscillation level near a stable equilibrium. Nonlinear modal interaction, however, can reverse this trend in multimachine systems. This letter gives a necessary-and-sufficient condition for strict leading-order local hardening, measured with respect to canonical action, in two symmetric three-machine operating-point families with equal or opposite power injections at the equal-inertia machines. An exact two-coordinate reduction expresses the condition in terms of a squared modal-frequency ratio and a dimensionless nonlinear coupling measure. The proof retains the induced mean shift and second harmonic and shows that the companion modal family remains locally softening. The criterion recovers both an equal-inertia hardening example and an exact large-inertia threshold. Coherent aggregation recovers the two-machine softening law, whereas the infinite-bus limit retains two dynamic rotor angles and permits hardening. An exact coherent-group extension preserves the selected periodic family, frequency, and canonical action for every finite machine count greater than two. Thus, three finite-inertia machines are the minimum permitting local hardening in the stated undamped, lossless classical model.
\end{abstract}

\vspace{-0.3in}
\section{Introduction}
Frequency--amplitude (F--A) curves characterize how electromechanical
oscillation frequencies vary with amplitude beyond the small-signal
regime~\cite{wang2016fa,wang2017properties}. For the single-degree-of-freedom (1-DOF)  swing equation with sinusoidal coupling, the leading local frequency correction is
negative, giving the familiar softening behavior. Nonlinear normal modes,
understood here as local periodic families, provide a periodic-orbit
interpretation of amplitude-dependent frequencies~\cite{kerschen2009nnm}.

However, in multimachine systems, nonlinear interaction between oscillatory modes can significantly affect the system response
\cite{messina2003modal}. This letter determines exactly when that interaction produces local hardening in two specified symmetric
three-machine operating-point families. Although the potentially hardening linear mode differs between the equal- and opposite-injection cases, both satisfy the same criterion: the squared feedback-to-target modal-frequency ratio \(R\) must lie between the coupling-dependent threshold $4\frac{1+2\chi}{1+3\chi}$ and the \(2{:}1\) resonance value \(4\), where \(\chi\) is a dimensionless nonlinear-coupling measure. Two exact examples illustrate the criterion, and coherent aggregation separates the scalar two-machine model from a two-machine--infinite-bus model with two dynamic rotor angles. An exact coherent-group extension preserves the selected periodic family in larger finite systems. The result concerns local frequency variation, not a stability margin or a classification of arbitrary network loading. 

\section{Symmetric System and Hardening Criterion}
Consider the undamped, lossless classical swing equations
\begin{equation}
 M_i\ddot\delta_i=P_i-\sum_{k\ne i}K_{ik}\sin(\delta_i-\delta_k),
 \label{eq:model}
\end{equation}
where $M_i>0$, $K_{ik}=K_{ki}\ge0$, and $P_i$ is a constant signed net power
injection at an internal machine node. Assume balanced injections
satisfying $\sum_iP_i=0$. Set
$M_1=M_2=M$, $M_3=mM$, $K_{13}=K_{23}=K$, and $K_{12}=\rho K$,
where $M,K,m>0$ and $\rho\ge0$. The network remains connected at $\rho=0$ via machine 3.

Using $\delta_3$ as the angular reference, consider two operating-point families parameterized by $\theta\in(-\frac{\pi}{2},\frac{\pi}{2})$: operating point A has equilibrium $\delta^*=(\theta,\theta,0)$ and injections
$(K\sin\theta,K\sin\theta,-2K\sin\theta)$;
Operating point B has equilibrium  $\delta^*=(\theta,-\theta,0)$ and injections
$(P,-P,0)$, where $P=K(\sin\theta+\rho\sin2\theta)$.
Thus, A has equal injections at machines 1 and 2, balanced by machine 3, whereas B has equal-and-opposite injections at machines 1 and 2 with zero injection at machine 3.
Machine 3 remains dynamic at finite $m$, including at B with zero net
injection. These families have three continuous parameters
$(m,\rho,\theta)$ and discrete type A or B; $M,K$ are positive scales.
The analysis here is restricted to these two symmetric loading patterns and does not cover general asymmetric power injections.

Let $c=\cos\theta>0$, $\tau=\tan\theta$ and $b=1+\frac{2}{m}$.
Removing the decoupled uniform-angle motion leaves two relative coordinates $x=\frac{\delta_1-\delta_2}{2}$ and
$y=\frac{\delta_1+\delta_2}{2}-\delta_3$. Their relative kinetic energy and momenta are:
\begin{equation}
 \begin{aligned}
T_r&=M\dot{x}^2+\frac{M}{b}\dot{y}^2,\\
p_x&=\frac{\partial T_r}{\partial\dot x}=2M\dot x,
\quad
p_y=\frac{\partial T_r}{\partial\dot y}
=\frac{2M}{b}\dot y.
 \end{aligned}
\end{equation}
Obtain their exact Hamiltonian by Legendre transformation:
\begin{equation}
 \begin{aligned}
 H_r&=p_x\dot x+p_y\dot y-(T_r-U)
 =\frac{p_x^2+bp_y^2}{4M}+U,\\
 U&=-K\rho\cos2x-2K\cos x\cos y\\
  &\quad -(P_1-P_2)x-(P_1+P_2)y.
 \end{aligned}
 \label{eq:energy}
\end{equation}
For the eqilibria $(x^*,y^*)=(0,\theta)$ for A and $(\theta,0)$ for B, $\nabla^2 U^*$ has
\(U_{xy}^*=0, ~U_{xx}^*=2Kca_i~(i=\text{A, B}), ~ U_{yy}^*=2Kc>0
\)
where $a_A=1+\frac{2\rho}{c}$, and $a_B=1+\frac{2\rho\cos2\theta}{c}$. Linearize Euler-Lagrange equations \(2M\ddot x+\frac{\partial U}{\partial x}=0\) and \(\frac{2M}{b}\ddot y+\frac{\partial U}{\partial y}=0\) about the equilibria to obtain two squared modal frequencies
\[
\omega_x^2=\frac{U_{xx}^*}{2M}=\omega_0^2a_i~(i=\text{A or B}),\qquad
\omega_y^2=\frac{b\,U_{yy}^*}{2M}=\omega_0^2b.
\]
where \(\omega_0^2=\frac{Kc}{M}\). A is automatically a nondegenerate relative energy minimum for \(|\theta|<\frac{\pi}{2}\) and \(\rho\ge0\) because \(a_A=1+\frac{2\rho}{c}>0\), whereas B additionally requires \(a_B=1+\frac{2\rho\cos2\theta}{c}>0\). This condition allows \(\cos2\theta<0\): the \(1\!-\!2\) line may have negative incremental synchronizing stiffness if \(|\theta|>\frac{\pi}{4}\),
while the total reduced stiffness remains positive.

Let $\omega$ be the target linear angular frequency from $\omega_x$ and $\omega_y$ and $\nu$ be the other
frequency. The target is the out-of-phase $x$ mode of machines 1 and 2
at A, and their in-phase $y$ mode relative to machine 3 at B.
Canonical action parameterizes the oscillation level of the complete periodic orbit:
\begin{equation}
 \begin{gathered}
 I=\frac1{2\pi}\oint(p_x\,dx+p_y\,dy), \quad
 \Omega(I)=\omega+\alpha I+O(I^2),
 \end{gathered}
 \label{eq:action}
\end{equation}
where $\Omega=2\pi/T$ is its fundamental angular frequency.
Local hardening and softening are classified by the first nonzero frequency correction in \eqref{eq:action}: strict leading-order local hardening means $\alpha>0$ and softening means $\alpha<0$; at $\alpha=0$, higher orders decide the trend.

Let the target linear mode be mass-normalized; i.e. its eigenvector $v$ satisfies \(v^{T}M_rv=1\), where \(M_r=\operatorname{diag}(2M, 2M/b)\) is the reduced inertia matrix. 

In mass-normalized modal coordinates $q,z$, $p_q=\dot q$ and
$p_z=\dot z$. For target first-harmonic amplitude $\eta$,
$q=\eta\cos(\Omega t)+O(\eta^3)$ and $z=O(\eta^2)$, giving
\begin{equation}
    I=\frac{1}{2\pi}\oint(p_q\,dq+p_z\,dz)
 =\frac{\omega\eta^2}{2}+O(\eta^4).
\end{equation}
For a rotor-angle amplitude $A=\ell\eta+O(\eta^2)$ with
$\ell>0$, $\Omega(A)=\omega+\frac{\alpha\omega}{2\ell^2}A^2
+O(A^3)$. Thus, its initial F--A slope is zero and its leading
quadratic coefficient has the sign of $\alpha$.

 In these modal coordinates, let 
\(
C=C_{kjj}=\left.
\frac{\partial^3U}
{\partial q_k\,\partial q_j^2}
\right|_*
\)
denote the third-order coupling derivative between target mode \(j\) and a companion mode \(k\), and 
\(
D=D_{jjjj}
=\left.
\frac{\partial^4U}
{\partial q_j^4}
\right|_*
\) 
the fourth-order derivative along the target mode. Define $R=\frac{\nu^2}{\omega^2}$, $\chi=-\frac{C^2}{D\nu^2}$, $S=-\frac{D}{8\omega^2}$, and
$R_c(\chi)=4\frac{1+2\chi}{1+3\chi}$.
Table~\ref{tab:map} maps the physical parameters to these quantities.

\begin{table}[t]
\caption{Parameters of the selected local modal family}
\label{tab:map}
\centering
\footnotesize
\begin{tabular}{c|ccccc}
\hline
Case 
& $\omega^2/\omega_0^2$ 
& $\nu^2/\omega_0^2$ 
& $R$ 
& $\chi$ 
& $S$ \\
\hline
\shortstack{A: Equal injections\\Target $x$}
& $a_A$
& $b$
& $\frac{b}{a_A}$
& $\frac{\tau^2}{4a_A-3}$
& $\frac{4a_A-3}{16Ma_A}$
\\
\shortstack{B: Opposite injections\\Target $y$}
& $b$
& $a_B$
& $\frac{a_B}{b}$
& $\frac{\tau^2}{a_B}$
& $\frac{b}{16M}$
\\
\hline
\end{tabular}
\end{table}

\begin{theorem}
For either specified operating-point family, with $a_B>0$ in B,
let $\omega$ and $\nu$ denote the selected target and companion
frequencies. Assume $\nu\ne n\omega$ for $n=1,2,\ldots$. Since the reduced system is analytic and the equilibrium is a nondegenerate energy minimum, the conservative Lyapunov subcenter theorem~\cite{llave2019lsm} guarantees a local family of periodic oscillations emerging from the target mode, with
\[
\Omega(I)=\omega+\alpha I+O(I^2),
\]
where
\begin{equation}
 \begin{gathered}
 \alpha=S\frac{(1+3\chi)R-4(1+2\chi)}{4-R},\\
 \boxed{\alpha>0\quad\Longleftrightarrow\quad R_c(\chi)<R<4.}
 \end{gathered}
 \label{eq:criterion}
\end{equation}
On this nonresonant domain, $\alpha<0$ for $R<R_c$ or $R>4$,
and $R=R_c$ is the zero-coefficient boundary when admissible.
The value $R=4$ is an excluded 2:1 resonance.
\end{theorem}

\begin{IEEEproof}
Set $X=\sqrt{2M}(x-x^*)$ and $Y=\sqrt{\frac{2M}{b}}(y-y^*)$.
The potential is even in the target displacement ($X$ for A, $Y$ for B),
so the target self-cubic derivative $C_{jjj}=0$. For the target mode $j$
and companion mode $k$, differentiating~\eqref{eq:energy} gives
{\small
\[
(C,D)=
\begin{cases}
\left(
-\dfrac{K\sin\theta\sqrt b}{\sqrt2 M^{3/2}},
-\dfrac{Kc(4a_A-3)}{2M^2}
\right), & \mathrm{A},\\[5pt]
\left(
-\dfrac{K\sin\theta b}{\sqrt2 M^{3/2}},
-\dfrac{Kcb^2}{2M^2}
\right), & \mathrm{B}.
\end{cases}
\]
}
From Table~\ref{tab:map}, $D<0$, $S>0$, and $\chi\ge0$.
For target coordinate $q$ and feedback coordinate $z$, the relevant
nonlinear potential terms are $\frac12{Cq^2z}+\frac{1}{24}Dq^4$. Let $q=\eta\cos\Omega t+O(\eta^3)$.
The forced second-mode response is
\[
 z=-\frac{C\eta^2}{4}
 \left[\frac1{\nu^2}+\frac{\cos2\Omega t}{\nu^2-4\omega^2}\right]
 +O(\eta^4).
\]
Reflection symmetry and local phase-fixed uniqueness make $q$ odd and
$z$ even in signed $\eta$. Other allowed potential terms do not affect
the leading correction because $q=O(\eta)$ and $z=O(\eta^2)$.
Canceling the third-order fundamental term and using~\eqref{eq:action}
gives
\begin{equation}
 \alpha=\frac{D}{8\omega^2}
 -\frac{C^2}{4\omega^2\nu^2}
 -\frac{C^2}{8\omega^2(\nu^2-4\omega^2)}.
 \label{eq:feedback}
\end{equation}
These are the direct quartic, mean-shift, and second-harmonic
contributions. For $C\ne0$, the nonlinear orbit includes an induced
second-mode response, even though its tangent is a linear mode.
Substitution of $R,\chi,S$ gives~\eqref{eq:criterion}.
Its numerator has the single zero $R_c\le4$; comparison with the
denominator $4-R$ proves both directions of the sign criterion.
\end{IEEEproof}

If \(\chi>0\), then \(\frac{8}{3}<R_c<4\); hence the broader interval \(\frac{8}{3}<R<4\) is necessary but not sufficient for hardening. Moreover, hardening implies
\(
\sqrt{\frac{8}{3}}<\frac{\nu}{\omega}<2,
\)
so the target-mode nonresonance condition is automatically satisfied. If \(\chi=0\), then \(R_c=4\) and \(\alpha=-S<0\) away from the excluded resonances. No resonance-independent finite amplitude neighborhood is implied as \(R\to4\).

\emph{Companion modes:} At A, $x=p_x=0$ is invariant and
$\alpha_y^A=-b\frac{3+5\tau^2}{48M}<0$.
At B, $y=p_y=0$ is invariant, with a scalar potential containing both
$\cos x$ and $\cos2x$. Its coefficient is
\[
 \alpha_x^B=-\frac{4a_B-3}{16Ma_B}
 -\frac{5\tau^2(1+8\rho c)^2}{48Ma_B^2}.
\]
To prove its sign even when $\cos2\theta<0$, set $v=c^2\in(0,1]$,
$u=\frac{\rho}{c}\ge0$. Direct expansion gives
$48Ma_B^2v\alpha_x^B=-\mathcal P$, where
$\mathcal P=(5-2v)(1+10uv)+16u^2v[3+8v(1-v)]>0$.
Thus, Theorem 1 determines whether the selected target family—\(x\) in A or \(y\) in B—hardens; the companion family is locally softening in both cases. Additional resonant branches are outside the claim of Theorem 1.

\vspace{-0.1in}
\section{Special Cases and Extensions}
\subsection{Equal-inertia hardening example} 
In A, choose $M=1$, $K=2$, $m=1$,
$\rho=\frac{1}{128}$, and $\theta=\frac{\pi}{3}$. The coupling is
$K_{12}=\frac{1}{64}$, and the injections are
$(\sqrt3,\sqrt3,-2\sqrt3)$. Then $a_A=\frac{33}{32}$, $b=3$,
$R=\frac{32}{11}$, $\chi=\frac{8}{3}$, and $R_c=\frac{76}{27}$, giving
$\alpha_A=\frac{7}{132}>0$ by~\eqref{eq:criterion}.

\vspace{-0.1in}
\subsection{Large-inertia case} 
For B, choose \(M=1\), \(K=\frac{\sqrt7}{2}\),
\(\rho=\frac{77}{4\sqrt7}\), and \(\theta=\arctan\!\left(\frac{\sqrt3}{2}\right)\).
Then \(K_{12}=\frac{77}{8}\), the injections are \((6\sqrt3,-6\sqrt3,0)\),
and \(Kc=1\), \(a_B=\frac{15}{4}\), \(\tau^2=\frac{3}{4}\), \(\chi=\frac{1}{5}\),
\(R_c=\frac{7}{2}\), and \(R=\frac{15m}{4(m+2)}\). Consequently,
\begin{equation}
 \alpha_B(m)=\frac{(m+2)(m-28)}{10m(m+32)},
 \label{eq:large}
\end{equation}
so hardening occurs exactly for $m>28$ on the stated local-family
domain. At $m=28$ only the leading
coefficient vanishes. For fixed $\rho,\theta$, $R$ decreases with $m$
in A but increases in B, while $\chi$ is fixed. Large inertia is
therefore not a universal requirement for hardening.

\vspace{-0.1in}
\subsection{Coherent two-machine reduction} 
For operating-point family A, the exact invariant subspace \(x=p_x=0\) combines machines 1 and 2 into an equivalent machine of inertia \(M_A=2M\), coupled by \(2K\) to machine 3 with inertia \(M_B=mM\). With reduced inertia
\(
\mu=\frac{M_AM_B}{M_A+M_B}=\frac{2M}{b},
\)
the relative motion satisfies
\(
\mu\ddot y=2K(\sin\theta-\sin y).
\)
The scalar mass-normalized formula
\(
\alpha_2=\frac{D_s}{8\omega_s^2}-\frac{5C_s^2}{24\omega_s^4},
\)
with \(\omega_s^2=\frac{2Kc}{\mu}\), \(C_s=-\frac{2K\sin\theta}{\mu^{3/2}}\), and \(D_s=-\frac{2Kc}{\mu^2}\), gives
\(
\alpha_2=-\frac{3+5\tau^2}{24\mu}<0.
\)
This holds for any two positive finite inertias. At fixed A injections, \(\rho\to\infty\) gives a rigid-link limit for the coherent slow motion, although undamped fast oscillations may persist; removing the \(1\!-\!2\) line does not give this reduction. Thus, one finite-inertia machine has no relative mode, two-machine motion is locally softening, and three machines can harden.

\vspace{-0.1in}
\subsection{Two-machine-infinite-bus limit} 
Let \(m\to\infty\), so \(b\to1\). The reduced inertia matrix \(\operatorname{diag}(2M,\frac{2M}{b})\) converges regularly to \(\operatorname{diag}(2M,2M)\), while the potential remains unchanged. In a frame rotating with its limiting constant speed, machine 3 becomes the fixed angular reference, leaving two dynamic rotor angles. At stable, nonresonant limiting points, \eqref{eq:criterion} therefore applies with \(b=1\). Case A cannot harden because \(R=\frac{1}{a_A}\le1\). Case B hardens exactly when \(R_c(\chi)<a_B<4\), with \(a_B=4\) excluded by the \(2{:}1\) resonance. For \eqref{eq:large}, \(a_B=\frac{15}{4}\) gives the nonresonant limit \(\alpha_\infty=\frac{1}{10}\).

\vspace{-0.1in}
\subsection{Exact finite-system extension} 
Replace machine 3 by a coherent group of \(r=N-2\) machines with
indices \(k=3,\ldots,N\), distribute its original power $P_3^{(0)}$ evenly among $r$ machines, while retaining machines 1 and 2 and their
mutual coupling:
{\small
\[
M_k=\frac{mM}{r},~P_k=\frac{P_3^{(0)}}{r},~K_{1k}=K_{2k}=\frac{K}{r},~K_{kj}=\frac{J}{r^2},~k\ne j,
\]
}where $k,j\in\{3,\ldots,N\}$. 
If all group members have the same rotor angle \(\delta_g\) and speed,
they remain coherent because their internal coupling torques vanish;
after division by inertia, each member obeys the same equation as the
original machine 3. Their aggregate inertia, injection, and coupling to
machines 1 and 2 are therefore exactly those of machine 3. Moreover,
\(
\sum_{k=3}^{N}p_k\,d\delta_k
=mM\dot{\delta}_g\,d\delta_g.
\)
The coherent replacement preserves the total inertia and the momentum–angle integral defining the action. Hence the selected periodic family has the same period, action, and \(\alpha\), with no factor \(r\).

so the coherent periodic family preserves its period, action, and hence
\(\alpha\), with no factor \(r\).

Perturbations that sum to zero within the coherent group give \(r-1\)
internal modes with squared frequency
\(
\frac{2Kc+J}{mM}.
\)
At a hardening operating point, choose \(J=mM\nu^2-2Kc\), giving
\(J=mKc>0\) for A and \(J=Kc(ma_B-2)>0\) for B. The internal modes then
have frequency \(\nu\), so the full relative spectrum consists of the
simple target frequency \(\omega\) and \(r\) modes at \(\nu\). Since
hardening requires \(\omega<\nu<2\omega\), the target-mode
nonresonance condition remains satisfied; multiplicity of the
companion frequency does not violate the conservative existence
theorem~\cite{llave2019lsm}. At other nonresonant operating points,
\(J>0\) can be chosen to avoid the countable target-resonance values.
Thus, whenever the spectral assumptions of Theorem~1 hold, the
coherent-group family satisfies the same hardening criterion for every
finite \(N\ge3\). No claim is made about the additional internal-mode
families.


\section{Conclusion}
A single necessary-and-sufficient criterion is derived for strict leading-order local hardening in the two symmetric three-machine families. Second-harmonic feedback can overcome direct and mean-shift softening. Coherent two-machine motion remains scalar and softening, while the two-machine--infinite-bus limit can retain hardening. The coherent-group extension preserves the selected family for any finite \(N\ge3\). Thus, three finite-inertia machines are the minimum permitting local hardening. These results are local and do not imply global monotonicity or finite-amplitude stability.

\section*{Acknowledgment}
OpenAI Codex \cite{codex} assisted with mathematical checks, identifying examples and special cases, and editing Sections II--III. The author retains responsibility for the submitted content.
\bibliographystyle{IEEEtran}
\bibliography{references}

@article{wang2016fa,
  author  = {Wang, Bin and Sun, Kai},
  title   = {Formulation and Characterization of Power System Electromechanical Oscillations},
  journal = {IEEE Trans. Power Syst.},
  year    = {2016},
  volume  = {31},
  number  = {6},
  pages   = {5082--5093},
  doi     = {10.1109/TPWRS.2016.2535384}
}

@article{wang2017properties,
  author  = {Wang, Bin and Su, Xiaowen and Sun, Kai},
  title   = {Properties of the Frequency-Amplitude Curve},
  journal = {IEEE Trans. Power Syst.},
  year    = {2017},
  volume  = {32},
  number  = {1},
  pages   = {826--827},
  doi     = {10.1109/TPWRS.2016.2553583}
}

@article{kerschen2009nnm,
  author  = {Kerschen, Ga{\"e}tan and Peeters, Maxime and Golinval, Jean-Claude and Vakakis, Alexander F.},
  title   = {Nonlinear Normal Modes, Part {I}: A Useful Framework for the Structural Dynamicist},
  journal = {Mech. Syst. Signal Process.},
  year    = {2009},
  volume  = {23},
  number  = {1},
  pages   = {170--194},
  doi     = {10.1016/j.ymssp.2008.04.002}
}

@article{llave2019lsm,
  author  = {{de la Llave}, Rafael and Kogelbauer, Florian},
  title   = {Global Persistence of {Lyapunov} Subcenter Manifolds as Spectral Submanifolds under Dissipative Perturbations},
  journal = {SIAM Journal on Applied Dynamical Systems},
  year    = {2019},
  volume  = {18},
  number  = {4},
  pages   = {2099--2142},
  doi     = {10.1137/18M1210344}
}

@misc{codex,
  author       = {{OpenAI}},
  title        = {{Codex}},
  year         = {2026},
  howpublished = {AI assistance system},
  url          = {https://openai.com/codex/},
  note         = {Accessed: Oct. 15, 2026}
}

@article{messina2003modal,
  author  = {A. R. Messina and E. Barocio},
  title   = {Assessment of Non-Linear Modal Interaction in Stressed
             Power Networks Using the Method of Normal Forms},
  journal = {Int. J. Electr. Power Energy Syst.},
  volume  = {25},
  number  = {1},
  pages   = {21--29},
  year    = {2003},
  doi     = {10.1016/S0142-0615(02)00015-7}
}
\end{document}